\documentclass[print, 
               notlongauthorlist 
              ]{nsr}

\volume{00}

\artnum{00}

\firstpage{1}

\datesubmitted{XX Month Year}
\datereceived{XX Month Year} 
\daterevised{XX Month Year}
\dateaccepted{XX Month Year} 
\datepublished{XX Month Year}

\doinum{doi/number}

\copyrightyear{2024}

\author[1]{Stephan I. Tzenov \footnote{Corresponding author e-mail address: \tt tzenov@jinr.ru.}$^{\orcidlink{0000-0001-8672-308X}}$}

\affil[1]{Veksler and Baldin Laboratory for High Energy Physics, Joint Institute for Nuclear Research, 6 Joliot-Curie Street, Dubna, Moscow Region, Russian Federation, 141980}

\runauth{Stephan I. Tzenov... }

\title{Frobenius-Perron Operator Approach to the Beam-Beam Interaction in Circular Colliders}

\begin{document}


\maketitle

\begin{abstract}
\noindent
Unlike most publications devoted to the application of the self-consistent method of the nonlinear Vlasov-Poisson system to the study of beam-beam interaction, in this article an alternative strategy using the elegant approach of the Frobenius-Perron operator for symplectic twist maps has been developed. A detailed analysis of the establishment of an equilibrium density distribution in phase space, as well as the behavior of the perturbed distribution function with respect to the coherent stability of the two beams, has been carried out. 

Using the Renormalization Group technique for the reduction of the Frobenius-Perron operator, the case where the unperturbed rotation frequency (unperturbed betatron tune) of the map is far from any structural resonance driven by the beam-beam kick perturbation has been analyzed in detail. It has been shown that up to second order in the beam-beam parameter, the renormalized map propagator with nonlinear stabilization describes a random walk of the angle variable, implying that there exists an equilibrium distribution function depending only on the action variable.  

The linearized Frobenius-Perron operators for each beam imply a discrete form of the linearized Vlasov equations, which essentially comprises a new method for calculating coherent beam-beam instabilities using a matrix mapping technique. In the special case of an isolated coherent beam-beam resonance, a stability criterion for coherent beam-beam resonances has been found in closed form. 

An intriguing particular concerning the effect of repeated beam-beam collisions on collider luminosity has been derived explicitly. An addition of luminosity per kick (small though, of the order of the beam-beam parameter) in the course of successive beam-beam collisions could be achieved. 
\end{abstract}



\section{\label{sec:intro}Introduction}

Charged particle beams in accelerators and storage rings are subjected to external forces that are often rapidly oscillating, such as conventional quadrupole focusing forces, radio-frequency accelerating fields, etc. In addition, collective self-consistent excitation fields can also be rapidly oscillating. A typical example is a ring collider device for the storage of subsequently colliding beams, where the evolution of each of the two beams is strongly influenced by the electromagnetic forces generated by the counter-propagating beam. The kick experienced by each beam is strictly localized only in a small region around the interaction point and is periodic with a period of one revolution along the machine circumference. 

The non-trivial problems and issues associated with the now-common term {\it beam-beam interaction} are quite longstanding. The first machines to start operating in collider mode in the distant 1965 were the electron-electron colliders VEP-1 at the Institute of Nuclear Physics (INP) in Novosibirsk and PSEC at the Stanford Linear Accelerator Center (SLAC), respectively. In the same year the electron-positron collider AdA at the National Laboratory in Frascati and five years later in 1970 the first hadron collider Intersecting Storage Rings (ISR) at CERN became operational. Even in the initial stage of the operation of accelerators in the collision mode of the stored beams, it was realized that beam-beam interaction significantly limits the luminosity - one of the main parameters of modern colliders. At the present time, it can be without hesitation stated that beam-beam interaction represents one of the most complex problems in the physics of accelerators and charged particle beams. Despite significant progress in understanding the relevant issues and underlying processes, there is still no comprehensive picture that encompasses all the features and physical details of beam-beam interaction. 

Active work on the development and construction of charged particle colliders began simultaneously in the late 1950s in the laboratories of Frascati (Italy), SLAC (USA) and the Institute of Nuclear Physics (former USSR). The first to operate was the electron-positron collider AdA, built under the direction of the Austrian theoretician Bruno Touschek in Frascati. However, the first results were published a year later (1966) than the observations of elastic scattering of electrons (1965) at the Soviet VEP-1 (Opposing Electron Beams), created under the direction of G. I. Budker and A. N. Skrinsky. A little later, colliding beams were obtained in the PSEC (Princeton-Stanford Experiment Collider). These first three colliders were test ones, demonstrating the possibility of studying the physics of elementary particles. The first hadron collider was the proton synchrotron ISR, launched in 1970 at CERN with beam energy of 32 GeV. The only linear collider in history is the SLC (Stanford Linear Collider), which operated from 1988 to 1998.

Without claiming complete exhaustiveness, here we shall try to trace the main and most important particulars and achievements in the theoretical description of beam-beam interaction. It is fair to say that the progress in numerical simulation of beam-beam interaction is significantly greater than the achievements of the theoretical models proposed so far. The important developments in this direction remain outside our main goal here and therefore we will not give them the necessary attention in the subsequent exposition. Historically, the first theoretical model of the beam-beam interaction is the so-called weak-strong model, also known as the incoherent beam-beam interaction, to which a special workshop \cite{Herrera} has been dedicated. In this model, it is assumed that one of the beams is strong and rigid and does not undergo significant changes (practically unmodified) in the collision process, and its role is to act on the other beam (considered weak and mobile), the latter playing the role of a dynamic probe and indicator of the interaction. Over the years, a huge number of articles have been published, which are dedicated to various analytical aspects, as well as to numerical simulation of the weak-strong model of beam-beam interaction. In hadron colliders, the natural damping mechanism does not exist, which can lead to classical diffusion along a network of intersecting stochastic layers (the so-called Arnold diffusion) characteristic for nonlinear dynamical systems with many degrees of freedom. This phenomenon was first described in Ref. \citenum{chirikov}. 

The realistic model reflecting the collective nature of the interaction between the two moving in opposite directions beams is called the strong-strong or coherent beam-beam interaction model for short. In this model, the evolution of the two beams occurs synchronously, as the electromagnetic field created by each beam is influencing and modifying the other one at the interaction point. Based on our awareness of the existing literature, the coherent beam-beam interaction in one dimension was first theoretically studied by Chao and Ruth \cite{chao} by solving the linearized Vlasov-Poisson equations. Since the pioneering work of Chao and Ruth, numerous papers based on the self-consistent Vlasov technique have been published, among which it is necessary to note the article by Yokoya et al. \cite{yokoya} and, at the first place, that by Alexahin \cite{alexahin}. In addition to the above two articles, the literature abounds with a whole host of interesting and important works \cite{zenyokoya,pestrik,hirata,pestrik1}, all of which, for obvious reasons, are impossible to mention, moreover this is not our main goal here. Based on the macroscopic hydrodynamic approach the results regarding the linear mode coupling, also known as the coherent beam-beam resonance \cite{chao} have been generalized in Ref. \citenum{tzenovPAC}. Unlike the standard technique for solving the Vlasov-Poisson system of equations in terms of action-angle variables used in all references mentioned above, the approach used in Ref. \citenum{tzendav} is implemented in a "mixed" phase space (old coordinates and new canonical momenta). In this way, the form of the Poisson equation for the beam-beam potential(s) in Cartesian coordinates is preserved, which is significantly simpler to handle analytically on one hand, and more computationally efficient on the other. 

The local nature of beam-beam interaction is an excellent testbed for the application of the symplectic mappings approach, which is unfortunately less popular as compared to the Vlasov-Poisson technique \cite{dragt,hirata1} at the present moment of time. A new approach to beam-beam interaction in circular colliders, based on the symplectic twist map method with subsequent regularization of the one-turn beam-beam map, has been developed in Refs. \citenum{tzenovEPAC,tzenovARX}. Therein, a regularized symplectic beam-beam map has been proposed, which correctly describes the long-term asymptotic behavior of the original dynamical system. It has been shown that the regularized map possesses an integral of motion that can be calculated in any desired order. The invariant density in phase space (stationary distribution function) has been constructed as a generic function of the integral of motion and a coupled system of nonlinear functional equations has been obtained for the distributions of the two colliding beams.

To study the coherent beam-beam instability, the present work follows a similar strategy \cite{tzenovFP}. The difference as compared to Refs. \citenum{tzenovEPAC,tzenovARX} is that instead of tracking individual trajectories in phase space, a statistical mechanics approach is applied via a distribution function of an ensemble of trajectories. The {\it Frobenius-Perron operator} of the density (distribution) function in phase space, sometimes called the {\it Transfer Operator} of this function or the {\it phase-space density propagator}, provides a powerful tool for studying the dynamics of recurrent iterations of the distribution function itself. In other words, the Frobenius-Perron operator provides a rule to determine how the evolution of phase-space densities over repeated iterations of the one-turn map is accomplished.

The article is organized as follows. In the following Section \ref{sec:hamdescr}, we briefly discuss the Hamiltonian formalism with an application to beam-beam interaction and the main tool for its description arising from this formalism, namely the coupled system of Vlasov-Poisson equations. Section \ref{sec:frobper} is devoted to the description of the Frobenius-Perron operator method, as well as to an outline of some of its main features and properties. The reduction of the Frobenius-Perron operator in the non-resonant case by the Renormalization Group method is performed in Section \ref{sec:froperRG}, while technical details are presented in \ref{sec:appendixA}. Based on the Frobenius-Perron operator, the problem of coherent beam-beam resonances in one dimension is studied in detail in Section \ref{sec:linearFP}. The influence of beam-beam collisions on the collider luminosity is studied in Section \ref{sec:lumin}. Last but not least, the conclusions and discussions of the obtained results, as well as future perspectives and possible developments are presented in Section \ref{sec:conclude}.

\section{\label{sec:hamdescr}Hamiltonian Description of Beam-Beam Interaction}

The two-dimensional model of coherent beam-beam interaction in a plane transversal to the individual particle orbits in each ${\left( k = 1, 2 \right)}$ beam is described by the Hamiltonian \cite{tzenovBOOK}
\begin{equation}
{\mathcal{H}}_k = {\frac {R} {2}} {\left( p_x^2 + p_y^2 \right)} + {\frac {1} {2 R}} {\left( G_x^{(k)} x^2 + G_y^{(k)} y^2 \right)} + \delta_p {\left( \theta \right)} {\frac {R q_k} {E_{sk} \beta_{sk}^2}} {\left( \varphi_{3-k} - c \beta_{sk} A_{s(3-k)} \right)}. \label{HamiltonK}
\end{equation}
Here, ${\left( x, p_x, y, p_y \right)}$ is the canonical conjugate pair of transverse variables, $R$ is the mean collider radius, $G_{x,y}^{(k)}$ are the linear machine focusing strengths for each beam in the transverse directions and the propagation of the latter is measured in terms of the azimuth $\theta$ adopted as the independent variable. In addition, $q_k$ are the corresponding particle charges, $c$ is the speed of light in vacuum, and $E_{sk}$ and $\beta_{sk}$ are the energy and the relative velocity of the synchronous particle, respectively. The periodic delta-function $\delta_p {\left( \theta \right)}$ multiplying the third term on the right-hand-side of Eq. \eqref{HamiltonK} reflects the locality of the interaction between the two beams occurring in the vicinity of the interaction point. The electromagnetic scalar $\varphi_{3-k}$ and the longitudinal component of the vector $A_{s(3-k)}$ potential describing the field created by the opposing beam satisfy the following equations 
\begin{equation}
{\boldsymbol{\nabla}}_{\perp}^2 \varphi_{3-k} = - {\frac {q_{3-k} N_{3-k} \varrho_{3-k}} {\epsilon_0}}, \label{WaveEqPhiK}
\end{equation}
\begin{equation}
{\boldsymbol{\nabla}}_{\perp}^2 A_{s(3-k)} = - \mu_0 q_{3-k} N_{3-k} J_{s(3-k)}, \label{WaveEqAsK}
\end{equation}
\begin{equation}
{\boldsymbol{\nabla}}_{\perp}^2 = {\frac {\partial^2} {\partial x^2}} + {\frac {\partial^2} {\partial y^2}}, \nonumber
\end{equation}
in the ultra-relativistic limit. Here $\epsilon_0$ and $\mu_0$ are the electric permittivity and the magnetic permeability of free space, respectively, $N_k$ is the particle number density of each beam, $\varrho_k$ is the normalized (to unity) beam density, and $J_{sk}$ is the beam current in longitudinal direction. Since $J_{s(3-k)} = - c \beta_{s(3-k)} \varrho_{3-k}$, one obtains the obvious relation 
\begin{equation}
A_{s(3-k)} = - {\frac {\beta_{s(3-k)}} {c}} \varphi_{3-k}. \label{Relation}
\end{equation}

It is convenient to perform an appropriate scaling of the beam-beam potential according to the relation 
\begin{equation}
\varphi_k = {\frac {q_k N_k \beta_{1k}^{\ast}} {4 \pi \epsilon_0}} V_k, \label{Scaling}
\end{equation}
and introduce the normalized canonical variables 
\begin{equation}
x = q_1 {\sqrt{\beta_{1k}}}, \qquad \qquad p_x = {\frac {1} {\sqrt{\beta_{1k}}}} {\left( p_1 - \alpha_{1k} q_1 \right)}, \label{NormCanVar}
\end{equation}
with similar relations for the vertical canonical conjugate pair ${\left( y, p_y \right)}$, indexed by 2. The quantities $\alpha$-s and $\beta$-s are the standard Twiss parameters, while the starred $\beta^{\ast}$ implies the corresponding Twiss parameter at the interaction point. Taking into account Eqs. \eqref{Relation} -- \eqref{NormCanVar}, the Hamiltonian \eqref{HamiltonK} can be cast in a normal form as 
\begin{equation}
{\mathcal{H}}_k = {\frac {{\dot{\chi}}_{1k}} {2}} {\left( p_1^2 + q_1^2 \right)} + {\frac {{\dot{\chi}}_{2k}} {2}} {\left( p_2^2 + q_2^2 \right)} + \delta_p {\left( \theta \right)} \lambda_k V_{3-k}. \label{HamilNFK}
\end{equation}
where $\chi_{1,2k}$ are the corresponding phase advances and the "raised dot" ${\dot{w}}$ implies differentiation of the corresponding variable $w$ with respect to the independent azimuthal variable $\theta$. Moreover, 
\begin{equation}
\lambda_k = {\frac {R r_p Z_k Z_{3-k} N_{3-k} \beta_{1(3-k)}^{\ast}} {A_k \gamma_{sk}}} {\frac {1 + \beta_{sk} \beta_{s(3-k)}} {\beta_{sk}^2}}, \label{BBParam}
\end{equation}
is the so-called {\it beam-beam parameter}, where $r_p$ is the classical proton radius, $Z_k$ and $A_k$ are the charge state and the mass number of the particles in the $k$-th beam, respectively, and $\gamma_{sk}$ is the corresponding Lorentz factor. The normalized beam-beam potential $V_{3-k}$ satisfies the Poisson equation 
\begin{eqnarray}
{\left( {\frac {\partial^2} {\partial q_1^2}} + \kappa_{3-k} {\frac {\partial^2} {\partial q_2^2}} \right)} V_{3-k} = - 4 \pi \varrho_{3-k}, \nonumber 
\\ 
\kappa_{3-k} = {\frac {\beta_{1(3-k)}^{\ast}} {\beta_{2(3-k)}^{\ast}}}. \label{Poisson}
\end{eqnarray}

For the sake of simplicity and clarity, we shall consider in what follows the one-dimensional case in one of the transversal degrees of freedom. The particle distribution function $f_k {\left( q, p; \theta \right)}$ of each beam is a solution to the Vlasov equation 
\begin{equation}
{\frac  {\partial f_k} {\partial \theta}} + {\dot{\chi}}_k p {\frac  {\partial f_k} {\partial q}} - {\frac  {\partial {\mathcal{H}}_k} {\partial q}} {\frac  {\partial f_k} {\partial p}} = 0, \label{VlasovK}
\end{equation}
while the normalized beam density is expressed as 
\begin{equation}
\varrho_k {\left( q; \theta \right)} = \int \limits_{-\infty}^{\infty} {\rm d} p f_k {\left( q, p; \theta \right)}, \label{DensityK}
\end{equation}
The locality of the beam-beam interaction, or in other words, the representation of the normalized beam-beam potential of the opposing beam as a thin electromagnetic element, suggests a substantial simplification of the problem. It will be demonstrated in what follows that the beam-beam coupling between the two colliding beams can be treated to a certain extent exactly by means of an elegant technique involving transfer maps. 


\section{\label{sec:frobper}The Frobenius-Perron Operator for the Beam-Beam Map}

It is assumed that the general reader is not familiar with the method of the Frobenius-Perron operator. For the sake of self-consistency of the exposition, we follow closely Ref. \citenum{tzenovBOOK}, and briefly dwell here on some of its basics and general properties. Consider a continuous multi-dimensional finite degree-of-freedom dynamical system (not necessarily Hamiltonian) defined by a state vector ${\mathbf{x}} {\left( t \right)}$, where $t$ denotes the independent (time) variable. The evolution of the system is described by the set of coupled first-order differential equations 
\begin{equation}
{\frac {{\rm d} {\mathbf{x}}} {{\rm d} t}} = {\mathbf{F}} {\left( {\mathbf{x}}, {\boldsymbol{\lambda}}; t \right)}, \label{DynEquat}
\end{equation}
where ${\mathbf{F}} {\left( {\mathbf{x}}, {\boldsymbol{\lambda}}; t \right)}$ is a vector field and ${\boldsymbol{\lambda}}$ is a set of additional parameters. Next, define the distribution function $f {\left( {\mathbf{x}}; t \right)}$ characterizing the statistical properties of the dynamical system and satisfying the Liouville equation 
\begin{equation}
{\frac {\partial f {\left( {\mathbf{x}}; t \right)}} {\partial t}} + {\boldsymbol{\nabla}} \cdot {\left[ {\mathbf{F}} {\left( {\mathbf{x}}, {\boldsymbol{\lambda}}; t \right)} f {\left( {\mathbf{x}}; t \right)} \right]} = 0. \label{LiouvEquat}
\end{equation}
Let us write the formal solution of the set of dynamic equations \eqref{DynEquat} as
\begin{equation}
{\mathbf{x}} {\left( t \right)} = {\mathbf{X}} {\left( {\mathbf{x}}_0, {\boldsymbol{\lambda}}; t \right)}, \label{SolDynEq}
\end{equation}
where ${\mathbf{x}}_0 = {\mathbf{X}} {\left( t_0 \right)}$ is the initial condition at some initial "time" $t_0$. One can then verify in a straightforward manner that the solution of the Liouville equation \eqref{LiouvEquat} reads as 
\begin{equation}
f {\left( {\mathbf{x}}; t \right)} = \int {\rm d} {\mathbf{z}} \delta {\left[ {\mathbf{x}} - {\mathbf{X}} {\left( {\mathbf{z}}, {\boldsymbol{\lambda}}; t \right)} \right]} f_0 {\left( {\mathbf{z}} \right)}, \label{SolLiouvEq}
\end{equation}
where $f_0 {\left( {\mathbf{x}} \right)}$ is the initial distribution function. 

For one-dimensional maps of the form
\begin{equation}
x_{n+1} = F {\left( x_n, \lambda \right)}, \label{OneDimMap}
\end{equation}
equation \eqref{SolLiouvEq} can be written as
\begin{equation}
f_{n+1} {\left( x \right)} = {\widehat{\mathbf{U}}} f_n {\left( x \right)} = \int {\rm d} z \delta {\left[ x - F {\left( z, \lambda \right)} \right]} f_n {\left( z \right)}, \label{SolMapEq}
\end{equation}
where ${\widehat{\mathbf{U}}}$ is the Frobenius-Perron operator. The defined above Frobenius-Perron operator can be written in a more explicit form as 
\begin{equation}
{\widehat{\mathbf{U}}} f_n {\left( x \right)} = \sum \limits_b {\frac {f_n {\left[ F_b^{-1} {\left( x, \lambda \right)} \right]}} {{\left| F^{\prime} {\left[ F_b^{-1} {\left( x, \lambda \right)} \right]} \right|}}}, \label{FrobPerOper}
\end{equation}
where the index $b$ runs over all the various branches of the inverse map $F^{-1}$ and $F^{\prime}$ implies differentiation with respect to $x$. The generalization to multidimensional maps is straightforward.

The iterative one-dimensional beam-beam map can be derived by formally solving the Hamilton’s equations of motion
\begin{equation}
{\dot{q}} = {\dot{\chi}}_k p, \qquad \quad {\dot{p}} = - {\dot{\chi}}_k q - \lambda_k \delta_p {\left( \theta \right)} V_{3-k}^{\prime} {\left( q; \theta \right)}, \label{HamEqMot}
\end{equation}
where the "prime" denotes the partial derivative with respect to the coordinate $q$. The result is \cite{tzenovEPAC,tzenovARX} 
\begin{equation}
q_{n+1} = q_n \cos \omega_k + {\left[ p_n - \lambda_k V_{3-k}^{\prime} {\left( q_n \right)} \right]} \sin \omega_k, \label{BBMapq}
\end{equation}
\begin{equation}
p_{n+1} = - q_n \sin \omega_k + {\left[ p_n - \lambda_k V_{3-k}^{\prime} {\left( q_n \right)} \right]} \cos \omega_k, \label{BBMapp}
\end{equation}
\begin{equation}
\omega_k = 2 \pi \nu_k, \label{RotNumb}
\end{equation}
where $\nu_k$ is the betatron tune related to the $k$-th beam, and $\omega_k$ is the corresponding one-turn betatron phase advance. It is worth emphasizing here that the beam-beam map presented above is symplectic, which can be easily verified by direct explicit check of its Jacobian. According to Eq. \eqref{SolMapEq} the Frobenius-Perron operator can be written as 
\begin{equation}
f_k^{(n+1)} {\left( q, p \right)} = \int {\rm d} \xi {\rm d} \eta \delta {\left\{ q - \xi c_k - {\left[ \eta - \lambda_k V_{3-k}^{\prime} {\left( \xi \right)} \right]} s_k \right\}} \delta {\left\{ p + \xi s_k - {\left[ \eta - \lambda_k V_{3-k}^{\prime} {\left( \xi \right)} \right]} c_k \right\}} f_k^{(n)} {\left( \xi, \eta \right)}, \label{FrobperOpera}
\end{equation}
where $c_k = \cos \omega_k$ and $s_k = \sin \omega_k$, respectively. In order to perform the integration, we manipulate the allowed values of the arguments 
\begin{equation}
q - \xi c_k - {\left[ \eta - \lambda_k V_{3-k}^{\prime} {\left( \xi \right)} \right]} s_k = 0, \label{Argum1}
\end{equation}
\begin{equation}
p + \xi s_k - {\left[ \eta - \lambda_k V_{3-k}^{\prime} {\left( \xi \right)} \right]} c_k = 0, \label{Argum2}
\end{equation}
of the delta-functions as follows. Multiplying Eq. \eqref{Argum1} by $c_k$ and Eq. \eqref{Argum2} by $s_k$, and subtracting the resulting equations, we obtain 
\begin{equation}
q c_k - p s_k - \xi = 0. \label{Argume1}
\end{equation}
Similarly, multiplying Eq. \eqref{Argum1} by $s_k$ and Eq. \eqref{Argum2} by $c_k$, and summing up the resulting equations, we find 
\begin{equation}
q s_k + p c_k - \eta + \lambda_k V_{3-k}^{\prime} {\left( \xi \right)} = 0, \label{Argume2}
\end{equation}
Rewriting Eq. \eqref{FrobperOpera} as 
\begin{equation}
f_k^{(n+1)} {\left( q, p \right)} = \int {\rm d} \xi {\rm d} \eta \delta {\left( q c_k - p s_k - \xi  \right)} \delta {\left[ q s_k + p c_k + \lambda_k V_{3-k}^{\prime} {\left( \xi \right)} - \eta \right]} f_k^{(n)} {\left( \xi, \eta \right)}, \label{FrobperOperat}
\end{equation}
the above integral becomes trivial to be taken, and the final form of the Frobenius-Perron operator can be expressed as 
\begin{equation}
f_k^{(n+1)} {\left( q, p \right)} = f_k^{(n)} {\left[ Q, P + \lambda_k V_{3-k}^{\prime} {\left( Q \right)} \right]}, \label{FrobPerFinal}
\end{equation}
where 
\begin{equation}
\begin{pmatrix}
Q \\
P
\end{pmatrix} = {\mathcal{R}}_k^T
\begin{pmatrix}
q \\
p
\end{pmatrix}, \qquad \quad {\mathcal{R}}_k = 
\begin{pmatrix}
\cos \omega_k & \sin \omega_k \\
- \sin \omega_k & \cos \omega_k
\end{pmatrix}, \label{Rotation}
\end{equation}
and the superscript $T$ implies matrix transposition. Formally replacing the arguments ${\mathbf{z}} = {\left( q, p \right)}^T$ by ${\mathcal{R}}_k {\mathbf{z}}$, we cast Eq. \eqref{FrobPerFinal} in a more convenient for subsequent use form 
\begin{equation}
f_k^{(n+1)} {\left( {\mathcal{R}}_k {\mathbf{z}} \right)} = f_k^{(n)} {\left[ q, p + \lambda_k V_{3-k}^{\prime} {\left( q \right)} \right]}. \label{FrobPerFin}
\end{equation}
Introducing the formal small parameter $\epsilon$ accounting for the perturbative character of the normalized beam-beam interaction potential, and the action-angle variables
\begin{equation}
q = {\sqrt{2 J}} \cos a, \qquad \qquad p = - {\sqrt{2 J}} \sin a, \label{ActAng}
\end{equation}
with 
\begin{equation}
J = {\frac {1} {2}} {\left( q^2 + p^2 \right)}, \qquad \qquad a = - \arctan {\left( {\frac {p} {q}} \right)}, \label{ActAngI}
\end{equation}
we write the Frobenius-Perron operator represented by Eq. \eqref{FrobPerFin} in the form 
\begin{equation}
f_k^{(n+1)} {\left( a + \omega_k, J \right)} = f_k^{(n)} {\left[ q, p + \epsilon \lambda_k V_{3-k}^{\prime} {\left( q \right)} \right]}. \label{FrobPerBas}
\end{equation}

Here is the place to pay particular attention to the following; exactly the same considerations are valid for the counter-circulating beam, for which a similar Frobenius-Perron operator can be derived. Thus, Eq. \eqref{FrobPerBas} and a similar one for the opposing beam becomes the main starting point of our further analysis. Let us also note that Eq. \eqref{Argume1} and \eqref{Argume2} represent the components of the inverse vector function according to the expression on the right-hand side of Eq. \eqref{FrobPerOper}. Applying the latter taking into account the fact that the denominator in Eq. \eqref{FrobPerOper} is unity (the beam-beam map is symplectic), one arrives at the same result for the Frobenius-Perron operator as above. Some additional mathematical properties of the Frobenius-Perron operator are presented in \ref{sec:appendixB}. 


\section{\label{sec:froperRG}Renormalization Group Reduction of the Frobenius-Perron Operator}

Equation \eqref{FrobPerBas} can be written in alternative form as 
\begin{equation}
f_k^{(n+1)} {\left( a + \omega_k, J \right)} = \exp {\left[ \epsilon \lambda_k {\left( \partial_q V_{3-k} \right)} \partial_p \right]} f_k^{(n)} {\left( a, J \right)}, \label{FrobPerAlt}
\end{equation}
where for brevity the notations $\partial_{(q,p)} = \partial / \partial (q, p)$ have been introduced. Since the beam-beam potential $V_{3-k}$ does not depend on the momentum variable $p$, we can write 
\begin{equation}
{\widehat{\mathbf{L}}}_{3-k} = {\left( \partial_q V_{3-k} \right)} \partial_p - {\left( \partial_p V_{3-k} \right)} \partial_q = {\left( \partial_q V_{3-k} \right)} \partial_p, \label{LiouvOper}
\end{equation}
where ${\widehat{\mathbf{L}}}_{3-k}$ is the Liouvillian operator associated with $V_{3-k}$. Let us note that in action-angle variables the Liouvillian operator acquires the form 
\begin{equation}
{\widehat{\mathbf{L}}}_{3-k} = {\left( \partial_a V_{3-k} \right)} \partial_J - {\left( \partial_J V_{3-k} \right)} \partial_a. \label{LiouvOperAA}
\end{equation}
Thus, Eq. \eqref{FrobPerAlt} becomes
\begin{equation}
f_k^{(n+1)} {\left( a + \omega_k, J \right)} = \exp {\left[ \epsilon \lambda_k {\widehat{\mathbf{L}}}_{3-k} \right]} f_k^{(n)} {\left( a, J \right)}. \label{FrobPerLiouv}
\end{equation}

Let us premise for the time being that the beam-beam potential $V_k {\left( q \right)}$ is a known function of the coordinate displacement $q$. Here we will not put forward any further assumptions about the nature of the beam-beam potential as a function of the canonical variable $q$; the latter depends essentially on the normalized density $\varrho {\left( q \right)}$. It is straightforward to check that the Fourier image of the beam-beam potential ${\widetilde{V}}_k {\left( \lambda \right)}$, defined as 
\begin{equation}
V_k {\left( q \right)} = {\frac {1} {2 \pi}} \int \limits_{-\infty}^{\infty} {\rm d} \lambda {\widetilde{V}}_k {\left( \lambda \right)} {\rm e}^{i \lambda q}, \label{BBPotFour1}
\end{equation}
\begin{equation}
{\widetilde{V}}_k {\left( \lambda \right)} = \int \limits_{-\infty}^{\infty} {\rm d} q V_k {\left( q \right)} {\rm e}^{-i \lambda q}. \label{BBPotFour2}
\end{equation}
possesses the following symmetry property 
\begin{equation}
{\widetilde{V}}_k^{\ast} {\left( \lambda \right)} = {\widetilde{V}}_k {\left( - \lambda \right)}, \label{SymProp}
\end{equation}
where the asterisk implies complex conjugation. Note that Eq. \eqref{BBPotFour2} can be written as 
\begin{equation}
{\widetilde{V}}_k {\left( \lambda \right)} = {\frac {4 \pi} {\lambda^2}} \int \limits_{-\infty}^{\infty} {\rm d} q {\rm d} p f_k {\left( q, p \right)} {\rm e}^{-i \lambda q} = {\frac {4 \pi} {\lambda^2}} \int \limits_{0}^{\infty} {\rm d} J \int \limits_0^{2 \pi} {\rm d} a f_k {\left( a, J \right)} {\rm e}^{-i \lambda {\sqrt{2 J}} \cos a}. \label{BBPotFour3}
\end{equation}
Using the Jacobi-Anger expansion \cite{Ryzhik,Stegun} 
\begin{equation}
{\rm e}^{i z \cos \varphi} = \sum \limits_{m=-\infty}^{\infty} i^m {\mathcal{J}}_m {\left( z \right)} {\rm e}^{i m \varphi}, \label{JacobiAng}
\end{equation}
where ${\mathcal{J}}_m {\left( z \right)}$ is the Bessel function of the first kind of order $m$, we represent the beam-beam potential in a Fourier series in the angle variable as follows 
\begin{equation}
V_k {\left( a, J \right)} = V_k^{(0)} {\left( J \right)} + V_k^{(a)} {\left( a, J \right)} = V_k^{(0)} {\left( J \right)} + 2 \sum \limits_{m=1}^{\infty} V_k^{(m)} {\left( J \right)} \cos {\left( m a \right)}. \label{BBPotFourier}
\end{equation}
The corresponding Fourier amplitudes are expressed in explicit form as 
\begin{equation}
V_k^{(0)} {\left( J \right)} = {\frac {1} {2 \pi}} \int \limits_{-\infty}^{\infty} {\rm d} \lambda {\widetilde{V}}_k {\left( \lambda \right)} {\mathcal{J}}_0 {\left( \lambda {\sqrt{2 J}} \right)}, \label{BBPotVk0}
\end{equation}
\begin{equation}
V_k^{(m)} {\left( J \right)} = {\frac {i^m} {2 \pi}} \int \limits_{-\infty}^{\infty} {\rm d} \lambda {\widetilde{V}}_k {\left( \lambda \right)} {\mathcal{J}}_m {\left( \lambda {\sqrt{2 J}} \right)}. \label{BBPotVkm}
\end{equation}
In obtaining the Fourier expansion given by Eq. \eqref{BBPotFourier}, the symmetry property $V_k^{(-m)} {\left( J \right)} = V_k^{(m)} {\left( J \right)}$ has been taken advantage of. 

Since the Frobenius-Perron operator approach is little known in the field of statistical description of nonlinear dynamical systems, it is worth devoting some attention, aside from the main exposition, to briefly acquainting the potential reader with its main characteristics and properties. Details can be found in \ref{sec:appendixA}, where the case when both rotation frequencies $\omega_k$ are far from nonlinear resonances excited by the beam-beam potentials $V_k$ is considered. For completeness, the resonance structure of the Frobenius-Perron operator can also be analyzed when one or both $\omega_k$ are relatively close to certain structural resonance(s) driven by the beam-beam potentials. The interested reader is referred to Ref. \citenum{tzenovFP} for details. It is instructive now to consider the non-resonant case, for which Eq. \eqref{FroPerRGeqcl} for the renormalized amplitude of the distribution function in the continuous limit can be written in the form 
\begin{equation} 
{\frac {\partial {\widetilde{F}}_k} {\partial n}} = - \omega_k \partial_a {\widetilde{F}}_k + {\left[ \lambda_k {\widehat{\mathbf{L}}}_{3-k}^{(0)} + \lambda_k^2 {\left( {\frac {1} {2}} {\widehat{\mathbf{L}}}_{3-k}^{(0) {\bf 2}} + \Omega_{3-k} \partial_a \right)} \right]} {\widetilde{F}}_k. \label{FokkerPlan} 
\end{equation}
Here [compare with Eqs. \eqref{LioOper0P} and \eqref{NonlinFreq}] 
\begin{equation}
{\widehat{\mathbf{L}}}_{3-k}^{(0)} = - \omega_{3-k}^{(u)} {\left( J \right)} \partial_a, \quad \quad  \omega_{3-k}^{(u)} {\left( J \right)} = \partial_J V_{3-k}^{(0)} {\left( J \right)}, \label{FoPlLieOper}
\end{equation}
is the nonlinear first-order incoherent tune-shift, while [in analogy with Eq. \eqref{FroPerNFreq}]  
\begin{equation}
\Omega_{3-k} {\left( \omega_k, J \right)} = \sum \limits_{m=1}^{\infty} m \cot {\left( {\frac {m \omega_k} {2}} \right)} \partial_J {\left( V_{3-k}^{(m)} \partial_J V_{3-k}^{(m)} \right)}, \label{FoPlaNFreq}
\end{equation}
is the nonlinear second-order incoherent tune-shift. Note that the operator ${\widehat{\mathbf{M}}}_{3-k}^{(0)}$ defined by Eq. \eqref{LioOper0PW} is neglected as providing a higher-order correction to the incoherent tune-shift.

Equation \eqref{FokkerPlan} exhibits a very important and far-reaching property - there exists an equilibrium solution for the renormalized distribution function ${\widetilde{F}}_k^{(0)} {\left( J \right)}$, which depends only on the action variables. Moreover, there exist a damping mechanism acting on the fluctuation harmonics with respect to the angle variables, such that the general solution of the Fokker-Planck equation \eqref{FokkerPlan} rapidly relaxes towards the invariant density distribution. 

The relaxation rate to the invariant density distribution depends on the first-order incoherent tune-shift, and for this reason we shall now turn to its calculation. Recalling that $V_k^{(0)} {\left( J \right)}$ is given by the expression \eqref{BBPotVk0}, and also that the Fourier image of the beam-beam potential is represented in the form of Eq. \eqref{BBPotFour3}, we obtain 
\begin{equation}
\omega_{3-k}^{(u)} {\left( J \right)} = - {\frac {4} {\sqrt{2 J}}} \int {\rm d} q {\rm d} p f_{3-k} {\left( q, p \right)}  \int \limits_0^{\infty} {\frac {{\rm d} \lambda} {\lambda}} {\mathcal{J}}_1 {\left( \lambda {\sqrt{2 J}} \right)} \cos {\left( q \lambda \right)}. \label{InCohTSInt}
\end{equation} 
The second integral (with respect to $\lambda$) in the above expression is tabular \cite{Prudnikov} and can be taken in a closed form as 
\begin{equation}
\int \limits_0^{\infty} {\frac {{\rm d} x} {x}} {\mathcal{J}}_n {\left( c x \right)} 
\begin{Bmatrix}
    \sin b x \\ 
    \cos b x
\end{Bmatrix} = {\frac {1} {n}} 
\begin{Bmatrix}
    \sin {\left[ n \arcsin {\left( b / c \right)} \right]} \\ 
    \cos {\left[ n \arcsin {\left( b / c \right)} \right]} 
\end{Bmatrix}, \label{Prudnik}
\end{equation}
for $0 < b \leq c$. Thus, we have 
\begin{equation}
\omega_{3-k}^{(u)} {\left( J \right)} = - {\frac {4} {\sqrt{2 J}}} \int \limits_{-\infty}^{\infty} {\rm d} p \int \limits_{-{\sqrt{2 J}}}^{\sqrt{2 J}} {\rm d} q f_{3-k} {\left( q, p \right)} \cos {\left[ \arcsin {\left( {\frac {q} {\sqrt{2 J}}} \right)} \right]}. \label{InCohTSInt1}
\end{equation} 
Taking into account the equilibrium distribution function 
\begin{equation}
f_{3-k}^{(0)} {\left( q, p \right)} = {\frac {1} {2 \pi \sigma_{3-k}^2}} \exp {\left( - {\frac {p^2 +q^2} {2 \sigma_{3-k}^2}} \right)}, \label{EquilDF}
\end{equation}
as well as the integral representation of the modified Bessel function ${\mathfrak{I}}_n {\left( z \right)}$ (see e.g. Ref. \citenum{Stegun}) 
\begin{equation}
{\mathfrak{I}}_n {\left( z \right)} = {\frac {1} {\pi}} \int \limits_0^{\pi} {\rm d} \tau {\rm e}^{z \cos \tau} \cos {\left( n \tau \right)}, \label{ModifBessel}
\end{equation}
we obtain 
\begin{equation}
\omega_{3-k}^{(u)} {\left( J \right)} = - {\frac {\sqrt{2 \pi}} {\sigma_{3-k}}} {\left[ {\mathfrak{I}}_0 {\left( {\frac {J} {2 \sigma_{3-k}^2}} \right)} + {\mathfrak{I}}_1 {\left( {\frac {J} {2 \sigma_{3-k}^2}} \right)} \right]} \exp {\left( - {\frac {J} {2 \sigma_{3-k}^2}} \right)}. \label{IncohTS}
\end{equation}
A similar expression for the incoherent beam-beam tune-shift has been previously obtained in Ref. \citenum{tzenovARX}.
\begin{figure}
\centering
\includegraphics[width=13cm]{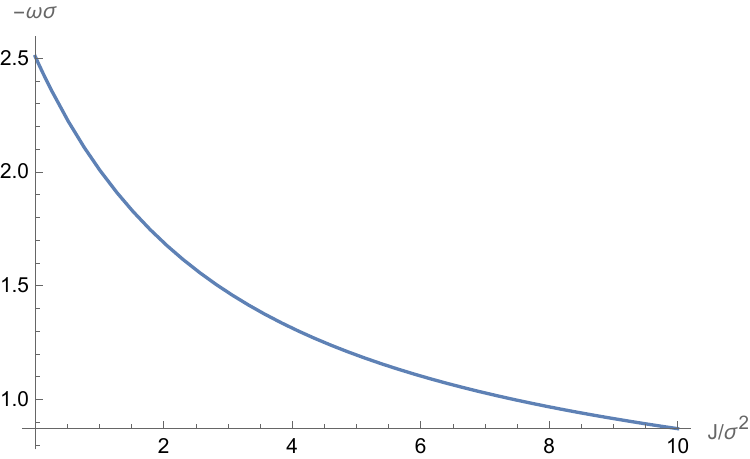}
\caption{\label{fig:tshift}Dependence of the first-order incoherent tune shift $- \omega_{3-k}^{(u)} \sigma_{3-k}$ as a function of the action variable $J / \sigma_{3-k}^2$ given by Eq. \eqref{IncohTS}.}
\end{figure} 

It is sometimes useful in practice to evaluate the averaged incoherent tune shift 
\begin{equation}
{\left \langle \omega_{3-k} \right \rangle} = \int \limits_0^{\infty} {\rm d} J f_{3-k}^{(0)} {\left( J \right)} \omega_{3-k}^{(u)} {\left( J \right)}. \label{AveIncohTS}
\end{equation}
Taking into account the expression \cite{Prudnikov}
\begin{equation}
\int \limits_0^{\infty} {\rm d} x {\rm e}^{- px} {\mathfrak{I}}_n {\left( cx \right)} = {\frac {c^n} {{\sqrt{p^2 - c^2}} {\left( p + {\sqrt{p^2 - c^2}} \right)}^n}}, \label{ModBessExpr}
\end{equation}
for $p > c$, we obtain 
\begin{equation}
{\left \langle \omega_{3-k} \right \rangle} = - {\frac {1} {2 \sigma_{3-k} {\sqrt{\pi}}}} {\frac {4 + 2 {\sqrt{2}}} {3 + 2 {\sqrt{2}}}}. \label{AverIncohTS}
\end{equation}
The dependence of the first-order incoherent tune shift as a function of the action variable is shown in Figure \ref{fig:tshift}. For typical characteristic parameters of the magnetic structure for the NICA collider at the Joint Institute for Nuclear Research (JINR) in Dubna and the number of particles in each of the beams $N_k \sim 4 \times 10^9$, the incoherent tune shift is of the order of $\lambda_k {\left \langle \omega_{3-k} \right \rangle} \sim 0.016$. 


\section{\label{sec:linearFP}Linearized Frobenius-Perron Operator and Stability of Coherent Beam-Beam Resonances}

In the preceding Section, it was shown that the equilibrium distribution function is an arbitrary function depending solely on the action variables. The answer to the question of what exactly and of what kind the invariant phase-space density should be anyway, was not given in an unambiguous form. To answer this question, a better strategy consists in the following; suppose that it is possible to construct integral(s) of motion [generalized action variable(s)] describing the dynamics of the beam-beam map \eqref{BBMapq} -- \eqref{BBMapp}. Then, the equilibrium distribution functions $G_{1,2}$ for the two beams satisfy coupled functional equations that can in principle be solved at least perturbatively. Such approach has been developed and followed in Ref. \citenum{tzenovARX}. In the present work we are not interested in the static equilibrium behavior of the beam distributions, but rather in the temporal evolution of the dynamic motions of the beam distributions around the equilibrium distributions $G_k$, that is 
\begin{equation}
f_k^{(n)} {\left( a, J \right)} = {\mathcal{F}}_k^{(n)} {\left( a, J \right)} + G_k {\left( J \right)}. \label{Linearizat}
\end{equation}
Substituting Eq. \eqref{Linearizat} into Eq. \eqref{FrobPerLiouv} and retaining only the first order terms in ${\mathcal{F}}_k^{(n)}$, we obtain the linearized Frobenius-Perron operator 
\begin{equation}
{\mathcal{F}}_k^{(n+1)} {\left( a + \omega_k, J \right)} = {\mathcal{F}}_k^{(n)} {\left( a - \lambda_k \omega_{3-k}^{(u)}, J \right)} + \lambda_k {\left[ \partial_a {\mathcal{V}}_{3-k}^{(n)} {\left( a - \lambda_k \omega_{3-k}^{(u)}, J \right)}  \right]} G_k^{\prime} {\left( J \right)}, \label{LineFrobPer}
\end{equation}
for $k = 1, 2$, where $G_k^{\prime}$ denotes derivative with respect to the action variable $J$. In addition 
\begin{equation}
{\mathcal{V}}_k^{(n)} {\left( a, J \right)} = 2 \int \limits_{-\infty}^{\infty} {\frac {{\rm d} \lambda} {\lambda^2}} {\rm e}^{i \lambda {\sqrt{2 J}} \cos a} \int {\rm d} a^{\prime} {\rm d} J^{\prime} {\mathcal{F}}_k^{(n)} {\left( a^{\prime}, J^{\prime} \right)} {\rm e}^{-i \lambda {\sqrt{2 J^{\prime}}} \cos a^{\prime}}, \label{FrobPerVk}
\end{equation}
are the first-order beam-beam potentials calculated with the perturbed distribution function ${\mathcal{F}}_k^{(n)} {\left( a, J \right)}$. In order to solve the linear recurrence equation \eqref{LineFrobPer}, we note that the linearized distribution function ${\mathcal{F}}_k^{(n)} {\left( a, J \right)}$ may be represented as 
\begin{equation}
{\mathcal{F}}_k^{(n)} {\left( a, J \right)} = G_k {\left( J \right)} \sum \limits_{l=-\infty}^{\infty} g_k^{(l)} {\left( J, n \right)} {\rm e}^{i l a}. \label{Represent}
\end{equation}
To this end, $g_k^{(l)} {\left( J, n \right)}$ are yet unknown Fourier harmonics, which, as will become clear from the subsequent exposition, determine the linear stability of the linearized Frobenius-Perron operator.

Assuming the equilibrium distribution function $G_k$ to be of the form \eqref{EquilDF}, for small beam sizes $\sigma_k$, the following formal trick 
\begin{equation} \nonumber 
\begin{split}
G_k {\left( J \right)} G_{3-k} {\left( J^{\prime} \right)} = {\mathcal{C}}_k \exp {\left( - {\frac {J} {\sigma_k^2}} - {\frac {J^{\prime}} {\sigma_{3-k}^2}} \right)} = {\mathcal{C}}_k \exp {\left( - {\frac {J^{\prime}} {\sigma_{3-k}^2}} + {\frac {J^{\prime}} {\sigma_k^2}} - {\frac {2 {\sqrt{J J^{\prime}}}} {\sigma_k^2}} \right)} 
\\ 
\times \exp {\left[ - {\frac {{\left( {\sqrt{J}} - {\sqrt{J^{\prime}}} \right)}^2} {\sigma_k^2}} \right]} \sim \sigma_k {\sqrt{\pi}} G_k {\left( J \right)} G_{3-k} {\left( J^{\prime} \right)} \delta {\left( {\sqrt{J}} - {\sqrt{J^{\prime}}} \right)}, 
\end{split}
\end{equation}
is valid. The above expression can be symmetrized with respect to the sizes of both beams and as a result we finally obtain
\begin{equation}
G_k {\left( J \right)} G_{3-k} {\left( J^{\prime} \right)} = {\overline{\sigma}} {\sqrt{\pi}} G_k {\left( J \right)} G_{3-k} {\left( J^{\prime} \right)} \delta {\left( {\sqrt{J}} - {\sqrt{J^{\prime}}} \right)}, \label{FormalTrick}
\end{equation}
where ${\overline{\sigma}} = {\left( \sigma_1 + \sigma_2 \right)} / 2$. Next, we substitute the ansatz \eqref{Represent} into Eq. \eqref{FrobPerVk}. To perform the integral entering Eq. \eqref{FrobPerVk}, we use the Jacobi-Anger identity \eqref{JacobiAng}, as well as the well-known relation between the Bessel functions \cite{Stegun}
\begin{equation}
{\mathcal{J}}_{n-1} {\left( z \right)} + {\mathcal{J}}_{n+1} {\left( z \right)} = {\frac {2 n} {z}} {\mathcal{J}}_n {\left( z \right)}. \label{RelatBessel}
\end{equation}
Equating similar harmonics with respect to the angle variable (proportional to ${\rm e}^{i l a}$) in the linearized Frobenius-Perron operator \eqref{LineFrobPer}, we obtain 
\begin{equation} 
G_k {\left( J \right)} g_k^{(l)} {\left( n + 1 \right)} = {\rm e}^{-il {\left( \omega_k + \lambda_k \omega_{3-k}^{(u)} \right)}} G_k {\left( J \right)} {\left[ g_k^{(l)} {\left( n \right)} + 2 J \lambda_k {\frac {\overline{\sigma}} {\sigma_k^2}} {\sqrt{2 \pi}} G_{3-k} {\left( J \right)} \sum \limits_{m=-\infty}^{\infty} {\boldsymbol{\mathcal{M}}}_{lm} g_{3-k}^{(m)} {\left( n \right)} \right]}. \label{LinFrobPerBas} 
\end{equation}
The infinite matrix ${\boldsymbol{\mathcal{M}}}$ can be expressed as 
\begin{equation}
{\boldsymbol{\mathcal{M}}}_{lm} = 
\begin{cases}
    {\dfrac {32il} {{\left[ {\left( l+m \right)}^2 - 1 \right]} {\left[ {\left( l-m \right)}^2 - 1 \right]}}}, & \qquad \text{if } l+m = {\rm even}, \\
    0,          & \qquad \text{if } l+m = {\rm odd},
\end{cases} \label{InfMatrix}
\end{equation}
where use has been made of the tabular integral \cite{Prudnikov} 
\begin{equation}
\int \limits_0^{\infty} {\frac {{\rm d} z} {z}} {\mathcal{J}}_m {\left( c z \right)} {\mathcal{J}}_n {\left( c z \right)} = {\frac {2} {\pi {\left( n^2 - m^2 \right)}}} \sin {\frac {{\left( n-m \right)} \pi} {2}}. \label{TabularInt}
\end{equation}

If $g_k^{(l)} {\left( n \right)}$ does not depend on the action variable, Eq. \eqref{LinFrobPerBas} can be further simplified by integrating away the action variable from its both sides. This approximation however, is valid if and only if the perturbed betatron tunes $\omega_k + \lambda_k \omega_{3-k}^{(u)}$ do not depend on the action $J$, which obviously is not the case. The dependence on the action variable leads to an effect similar to Landau damping, well-known in plasma physics, which we shall neglect in what follows. Another justification for the validity of such an approximation is the rapid decrease of the incoherent tune shift as a functional dependence on the action variable $J$ clearly visible in Figure \ref{fig:tshift}. Thus, the first-order incoherent tune shift can be approximately replaced by its average value given by Eq. \eqref{AverIncohTS}. With all of the above in hand, Eq. \eqref{LinFrobPerBas} can be cast in the form of a single-turn map 
\begin{equation} 
g_k^{(l)} {\left( n + 1 \right)} = {\rm e}^{-il {\left( \omega_k + \lambda_k {\left \langle \omega_{3-k} \right \rangle} \right)}} {\left[ g_k^{(l)} {\left( n \right)} + {\widetilde{\lambda}}_k \sum \limits_{m=-\infty}^{\infty} {\boldsymbol{\mathcal{M}}}_{lm} g_{3-k}^{(m)} {\left( n \right)} \right]}, \label{FrobPerOTM} 
\end{equation}
where 
\begin{equation}
{\widetilde{\lambda}}_k = {\sqrt{\frac {2} {\pi}}} \lambda_k {\frac {{\overline{\sigma}} \sigma_{3-k}^2} {\Sigma^4}}, \qquad \qquad \Sigma^2 = \sigma_1^2 + \sigma_2^2, \label{TildBBPar}
\end{equation}
\begin{figure}
\centering
\includegraphics[width=13cm]{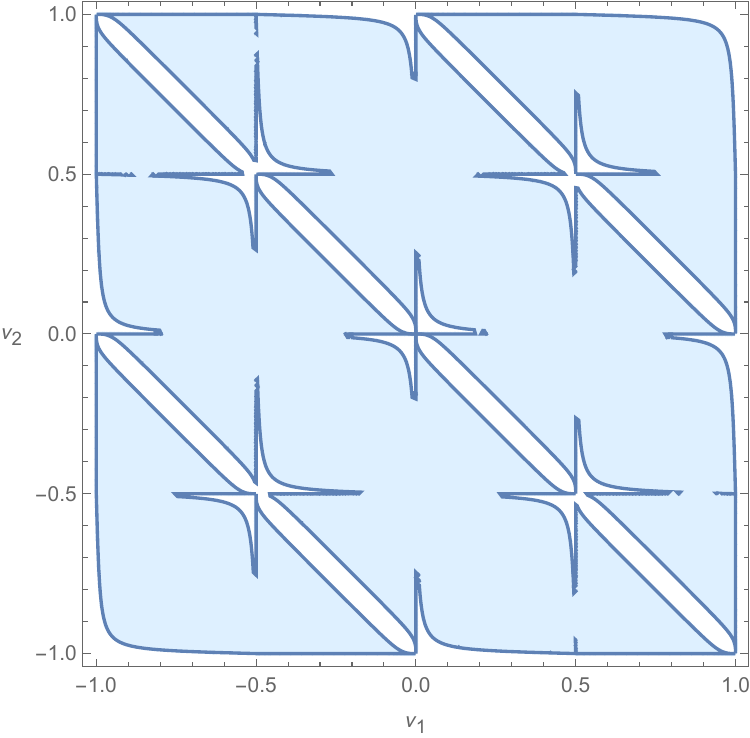}
\caption{\label{fig:stab1} Stability diagram (the shaded region) of a coherent beam-beam coupling resonance of the form given by Eq. \eqref{CohBBReson}, where $n_1 = n_2 = 1$. The plot is presented in the fractional part of the tune ${\left( \nu_1, \nu_2 \right)}$-space. For demonstrativeness, the beam-beam parameter is taken to be $\lambda_k \sim 4.7712 \times 10^{-5}$.}
\end{figure} 

Consider now an isolated coherent beam-beam resonance of the form 
\begin{equation}
n_1 {\widetilde{\omega}}_1 + n_2 {\widetilde{\omega}}_2 = 2 \pi s + \Delta, \label{CohBBReson}
\end{equation}
where $n_1$, $n_2$ and $s$ are integers, the quantity $\Delta$ is the resonance detuning, and 
\begin{equation}
{\widetilde{\omega}}_k = \omega_k + \lambda_k {\left \langle \omega_{3-k} \right \rangle}, \label{TildeTune}
\end{equation}
are the perturbed betatron tunes for each beam. To study the stability of the isolated coherent beam-beam resonance of the form \eqref{CohBBReson}, we retain only the $\pm n_1$ and the $\pm n_2$ elements in the infinite matrix ${\boldsymbol{\mathcal{M}}}_{lm}$, the transformation matrix of the coupled map equations \eqref{FrobPerOTM} can be expressed as 
\begin{equation}
\begin{pmatrix}
    {\rm e}^{-i \psi_1} & 0 & \alpha_1 {\rm e}^{-i \psi_1} & \alpha_1 {\rm e}^{-i \psi_1} \\ 
    0 & {\rm e}^{i \psi_1} & -\alpha_1 {\rm e}^{i \psi_1} & -\alpha_1 {\rm e}^{i \psi_1} \\ 
    \alpha_2 {\rm e}^{-i \psi_2} & \alpha_2 {\rm e}^{-i \psi_2} & {\rm e}^{-i \psi_2} & 0 \\ 
    -\alpha_2 {\rm e}^{i \psi_2} & -\alpha_2 {\rm e}^{i \psi_2} & 0 & {\rm e}^{i \psi_2} 
\end{pmatrix}
, \label{TransfMat}
\end{equation}
where 
\begin{equation} 
\psi_k = n_k {\widetilde{\omega}}_k, \qquad \qquad  \alpha_1 = {\widetilde{\lambda}}_1 {\boldsymbol{\mathcal{M}}}_{n_1n_2}, \qquad \qquad \alpha_2 = {\widetilde{\lambda}}_2 {\frac {n_2} {n_1}} {\boldsymbol{\mathcal{M}}}_{n_1n_2}. \label{Paramets} 
\end{equation}
The transition matrix \eqref{TransfMat} contains all the information about the stability of our system in the process of successive beam-beam kicks, therefore it is imperative to investigate its eigenvalues. The latter are the roots of the secular equation 
\begin{equation}
{\left( \mu^2 - 2 \mu \cos \psi_1 + 1 \right)} {\left( \mu^2 - 2 \mu \cos \psi_2 + 1 \right)} + 2 \alpha_1 \alpha_2 {\left[ \cos {\left( \psi_1 - \psi_2 \right)} - \cos {\left( \psi_1 + \psi_2 \right)}  \right]} \mu^2 = 0. \label{SeculEquat}
\end{equation}
The above equation \eqref{SeculEquat} can be converted into a more convenient form as follows 
\begin{equation} 
{\left( \mu^2 - 2 c_1 \mu  + 1 \right)} {\left( \mu^2 - 2 c_2 \mu + 1 \right)} = 0, \label{SecEquaConv}
\end{equation}
where 
\begin{equation} 
c_{1,2} = {\frac {1} {2}} {\left( \cos \psi_1 + \cos \psi_2 \right)} \pm {\frac {1} {2}} {\sqrt{ {\left( \cos \psi_1 - \cos \psi_2 \right)}^2 - 4 A \sin \psi_1 \sin \psi_2 }}, \label{Coeffc1c2} 
\end{equation}
\begin{equation}
A = {\widetilde{\lambda}}_1 {\widetilde{\lambda}}_2 {\frac {n_2} {n_1}} {\boldsymbol{\mathcal{M}}}_{n_1n_2}^2. \label{CoefficA}
\end{equation}
The motion is stable if the coefficients $c_{1,2}$ given by Eq. \eqref{Coeffc1c2} simultaneously satisfy the conditions 
\begin{equation}
-1 \leq c_{1,2} \leq 1. \label{StabilCond}
\end{equation}
\begin{figure}
\centering
\includegraphics[width=13cm]{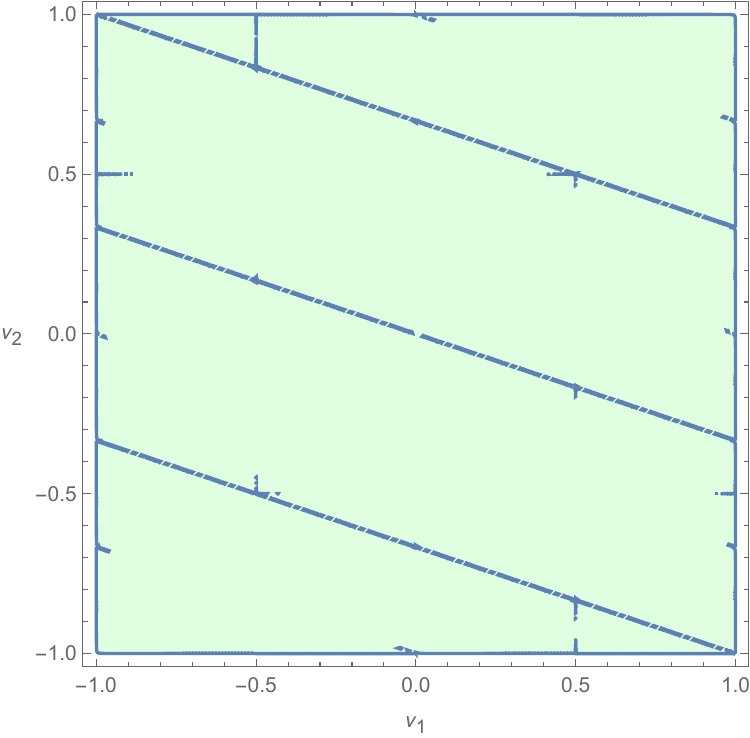}
\caption{\label{fig:stab2} Stability diagram (the shaded region) of a coherent nonlinear beam-beam resonance of the form given by Eq. \eqref{CohBBReson}, where $n_1 = 1$, $n_2 = 3$. The plot is presented in the fractional part of the tune ${\left( \nu_1, \nu_2 \right)}$-space. For demonstrativeness, the beam-beam parameter is taken to be $\lambda_k \sim 4.7712 \times 10^{-5}$.}
\end{figure} 

Figure \ref{fig:stab1} shows the stability region (shaded area) of the linear beam-beam coupling resonance ${\widetilde{\omega}}_1 + {\widetilde{\omega}}_2 = 2 \pi s + \Delta$ in the space of the fractional part of the shifted betatron tunes ${\widetilde{\nu}}_1$ and ${\widetilde{\nu}}_2$ [compare with Eq. \eqref{TildeTune}]. For a better clarity in the visualization of the structure and shape of the islands of instability, an increased value of the beam-beam parameter $\lambda_k$ corresponding to $N_k \sim 4 \times 10^{10}$ number of particles in each beam has been taken. 
\begin{figure}
\centering
\includegraphics[width=13cm]{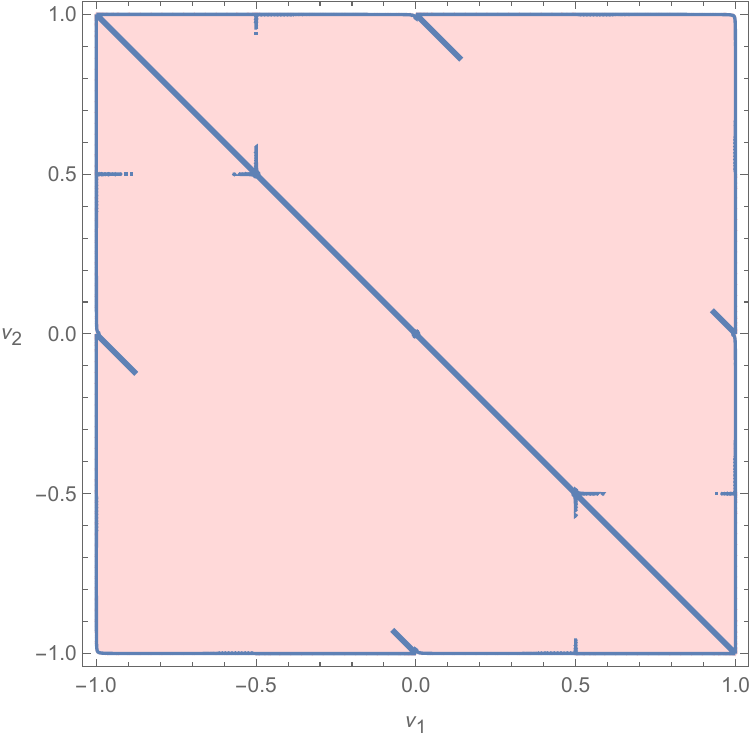}
\caption{\label{fig:stab3} Stability diagram (the shaded region) of a coherent beam-beam coupling resonance of the form given by Eq. \eqref{CohBBReson}, where $n_1 = n_2 = 1$. The plot is presented in the fractional part of the tune ${\left( \nu_1, \nu_2 \right)}$-space. The beam-beam parameter is taken to be $\lambda_k \sim 4.7712 \times 10^{-6}$, which corresponds to the realistic case, where $N_k \sim 4 \times 10^9$.}
\end{figure} 

Note that according to Eq. \eqref{InfMatrix} only nonlinear beam-beam resonances of even order are possible. Furthermore, the elements of the infinite matrix ${\mathcal{M}}_{lm}$ decrease quite rapidly with the resonance order, which leads to a drastic reduction of the resonant driving term. For comparison, Figure \ref{fig:stab2} shows the stability diagram in the case of a fourth-order coherent nonlinear beam-beam resonance ${\widetilde{\omega}}_1 + 3 {\widetilde{\omega}}_2 = 2 \pi s + \Delta$. The instability region consists of narrow resonance stopbands together with islands of instability scattered around them. There is a sufficiently wide band of stability, which greatly facilitates the felicitous selection of the operating betatron tunes. In this sense, nonlinear coherent beam-beam resonances are significantly less dangerous than the linear coupling resonance. 

Finally, Figure \ref{fig:stab3} presents the realistic situation showing the stability diagram of the linear coherent beam-beam resonance at a value of the beam-beam parameter $\lambda_k \sim 4.7712 \times 10^{-6}$ corresponding to $N_k \sim 4 \times 10^9$ number of particles in each beam. A central narrow resonance stopband and scattered satellite narrow stopbands and small islands of instability are clearly visible. 

Figures \ref{fig:stab1} -- \ref{fig:stab3} refer to typical characteristic parameters of the NICA collider magnetic and interaction point structure being under construction at the Joint Institute for Nuclear Research (JINR) for fully stripped gold atoms $^{197}{\rm Au}^{79+}$.

\section{\label{sec:lumin}Beam-Beam Collisions Effect on Collider Luminosity} 

In modern colliders, in addition to the energy of the circulating charged particle beams, the number of beneficial interactions (events) in the course of their successive collisions at the interaction point(s) is also of particular importance. The quantity that measures the ability of a particle accelerator to produce the required number of interactions is called the {\it luminosity}. In the case of head-on collisions, the luminosity is proportional to the  overlap integral of the colliding beams' density distribution functions \cite{meshkov,herr}, and is given by the expression 
\begin{equation}
{\mathcal{L}} = \kappa L, \qquad \qquad \kappa = 2 N_1 N_2 {\mathfrak{f}} N_b, \label{Lumin}
\end{equation}
where ${\mathfrak{f}}$ is the revolution frequency and $N_b$ is the number of bunches in one beam. The quantity $L$ is the overlap integral 
\begin{equation}
L = \int \limits_{-\infty}^{\infty} {\rm d} q \varrho_1 {\left( q \right)} \varrho_2 {\left( q \right)}, \label{LumOverlap}
\end{equation}
where the normalized beam density $\varrho_k {\left( q \right)}$ of each beam ${\left( k = 1, 2 \right)}$ is given by Eq. \eqref{DensityK}. 

It is interesting and important to estimate the luminosity variation per unit collision given an established equilibrium distribution. For this purpose it is necessary to represent the expansion of the Frobenius-Perron operator \eqref{FrobPerFinal} about the equilibrium distribution $G_k {\left( q, p \right)}$ up to first order in terms of the small formal (beam-beam) parameter as 
\begin{equation} 
f_k {\left( q, p \right)} = G_k {\left( q, p \right)} + \lambda_k {\left[ \partial_Q V_{3-k} {\left( Q \right)} \right]} \partial_P G_k {\left( q, p \right)}. \label{LumFirOrd} 
\end{equation}
Note that the orthogonal transformation \eqref{Rotation} from the old ${\left( q, p \right)}$ canonical variables to the new ones ${\left( Q, P \right)}$ does not change the overall appearance of the quadratic form entering the equilibrium distribution function \eqref{EquilDF}. The next step is to calculate the perturbed beam density in configuration space 
\begin{equation} 
\varrho_k {\left( q \right)} = \varrho_{0,k} {\left( q \right)} + \lambda_k \int \limits_{-\infty}^{\infty} {\rm d} p {\left[ \partial_Q V_{3-k} {\left( Q \right)} \right]} \partial_P G_k {\left( q, p \right)}. \label{LumBeamDens} 
\end{equation}
Integration by parts and utilization of the fact that the beam-beam potential satisfies the Poisson equation \eqref{Poisson}, yields 
\begin{equation} 
\varrho_k {\left( q \right)} = \varrho_{0,k} {\left( q \right)} - 2 \pi \lambda_k \sin 2 \omega_k \int \limits_{-\infty}^{\infty} {\rm d} p G_k {\left( q, p \right)} \varrho_{0,3-k} {\left( q \cos \omega_k - p \sin \omega_k \right)}. \label{LumBeaDens} 
\end{equation}
The well-known Gaussian integral 
\begin{equation} 
\int \limits_{-\infty}^{\infty} {\rm d} x \exp {\left[ - {\left( a x^2 + b x + c \right)} \right]} = {\sqrt{\frac {\pi} {a}}} \exp {\left( {\frac {b^2} {4 a}} - c \right)}, \label{LumGauss} 
\end{equation}
can be used to perform the integral in Eq. \eqref{LumBeaDens}. The result is 
\begin{equation} 
\varrho_k {\left( q \right)} = \varrho_{0,k} {\left( q \right)} - \Delta \varrho_k {\left( q \right)}, \label{LumBeaDenst} 
\end{equation}
where 
\begin{equation} 
\Delta \varrho_k {\left( q \right)} = {\frac {\lambda_k \sin 2 \omega_k} {\sigma_k {\cal S}_k}} \exp {\left[ - {\frac {q^2} {2 \sigma_{3-k}^2}} {\left( {\frac {{\cal C}_k^2} {\sigma_k^2}} - {\frac {\sigma_k^2 s_k^2 c_k^2} {{\cal S}_k^2}} \right)} \right]}. \label{LumBDelDens} 
\end{equation}
Here, as before $s_k = \sin \omega_k$ and $c_k = \cos \omega_k$, while the quantities ${\cal S}_k$ and ${\cal C}_k$ are given by the expressions 
\begin{equation} 
{\cal S}_k^2 = \sigma_k^2 s_k^2 + \sigma_{3-k}^2, \qquad \qquad {\cal C}_k^2 = \sigma_k^2 c_k^2 + \sigma_{3-k}^2. \label{LumCoeff} 
\end{equation}

Finally, the luminosity variation per unit collision can be expressed as 
\begin{equation} 
\Delta L = \Delta L_1 + \Delta L_2, \label{LumVarTot} 
\end{equation}
where 
\begin{equation}
\Delta L_1 = - \int \limits_{-\infty}^{\infty} {\rm d} q \varrho_{0,2} {\left( q \right)} \Delta \varrho_1 {\left( q \right)}, \label{LumVar1}
\end{equation}
and a similar expression for $\Delta L_2$, in which the indices "$1$" and "$2$" on the right-hand side swap places. More explicitly 
\begin{equation}
\Delta L_k = - {\frac {\lambda_k \sin 2 \omega_k} {\sigma_k {\cal S}_k}} {\left( 1 + {\frac {{\cal C}_k^2} {\sigma_k^2}} - {\frac {\sigma_k^2 s_k^2 c_k^2} {{\cal S}_k^2}} \right)}^{-1/2}. \label{LumVarExplic}
\end{equation}

There are several intriguing features in Eq. \eqref{LumVarExplic} striking at first glance that are worth commenting on briefly here. First, the luminosity variation per unit collision is proportional to the beam-beam parameter $\lambda_k$, which is to be expected. An even more important peculiarity is the dependence of $\Delta L_k$ on the phase advance per one revolution $\omega_k$. This essentially means that the betatron tunes $\nu_k$ can be chosen such that $\sin 2 \omega_k < 0$, which directly implies $1/4 < {\rm Frac} {\left( \nu_k \right)} < 1/2$, where ${\rm Frac} {\left( \nu_k \right)}$ being the fractional part of the betatron tune. In this way an addition of luminosity per kick (although of the order of the beam-beam parameter) in the course of successive beam-beam collisions could be achieved.

\section{\label{sec:conclude}Concluding Remarks} 

As already mentioned, the beam-beam phenomenon is a very difficult subject, which includes many different effects, subdivided on the one hand into incoherent and coherent effects, and on the other hand into equilibrium and non-equilibrium ones. The equilibrium ones include the process of establishing an equilibrium phase-space density distribution, and among the non-equilibrium processes, the nonlinear oscillations and resonance phenomena with possible transition to stochasticity should be mentioned. There also exist quasi-equilibrium processes induced by the beam-beam force, which qualify as bifurcation phenomena leading to the highly unfavorable effect, also known as the "flip-flop" effect \cite{tzendav}.

In the present work, a detailed analysis of the establishment of an equilibrium density distribution in phase space and the relaxation towards the latter has been studied analytically. Furthermore, the behavior of the perturbed from equilibrium distribution function with respect to the coherent stability of the colliding beams, is carried out in linear approximation. Although we do not claim that the latter is the only important effect that arises in the beam-beam interaction, we suggest that coherent effects are indeed one of the most dominant beam-beam features. Unlike most publications devoted to the application of the self-consistent method of the nonlinear Vlasov-Poisson system to the study of beam-beam interaction, in this paper we have chosen an alternative strategy using the elegant approach of the Frobenius-Perron operator for symplectic twist maps. 

The Renormalization Group (RG) method has been applied to study the stochastic properties of the Frobenius-Perron operator for symplectic twist maps of the most general type and in particular for the beam-beam twist map. After a brief introduction and derivation of the Frobenius-Perron operator for a beam-beam symplectic map with rotation, the case where the unperturbed rotation frequency (unperturbed betatron tune) of the map is far from the structural resonances driven by the beam-beam kick perturbation has been analyzed in detail. It has been shown that up to second order in the beam-beam perturbation kick, the renormalized map propagator (equivalently, the renormalized Frobenius-Perron operator) with nonlinear stabilization ${\left( {\widehat{\mathbf{L}}}_k^{(0)} \neq 0 \right)}$ describes a random walk of the angle variable. This in turn implies two important consequences: first, there exists an equilibrium distribution depending only on the action variable and second, the relaxation rate to this invariant distribution depends on the nonlinear (incoherent) tune shift and takes place only with respect to the angle variable. Further, the incoherent beam-beam tune shift as a function of the action variable has been calculated explicitly. 

The linearized Frobenius-Perron operator for each of the two beams actually implies a discrete form of the linearized Vlasov equations. This essentially is equivalent to and signifies a new method for calculating coherent beam-beam instabilities using a matrix mapping technique. It offers a very simple description of the coherent beam-beam interaction and allows straightforward numerical calculations. In particular, the handling of the infinite system of linear map equations (as far as this is practically possible) allows the simultaneous treatment of all nonlinear resonances, taking into account the coupling between them. In the special case of an isolated coherent beam-beam resonance, a stability criterion for coherent beam-beam resonances has been found in closed form. 

An intriguing particular concerning the effect of individual successive beam-beam collisions on collider luminosity has been found explicitly. An addition of luminosity per kick (small though, of the order of the beam-beam parameter) in the course of successive beam-beam collisions could be achieved. 

As for other merits of the method described here that have remained hidden in the main body of the article, it is worth noting two additional ones as follows. The Frobenius-Perron operator approach can be generalized without much difficulty to systems with more than one degree of freedom, so as to cover both transverse directions and, if necessary, the longitudinal degree of freedom as well. Combined with an adequate Poisson solver, the Frobenius-Perron operator, especially in its Cartesian coordinate and momentum representation, can represent a tool of particular value for the numerical simulation of the beam-beam interaction. Its numerical implementation may provide a wonderful opportunity not only to track the orbits of individual particles, but also to follow and describe the dynamic evolution of an entire statistical distribution of an ensemble of particles. Such generalizations can turn the Frobenius-Perron operator approach into an indispensable tool in studying beam-beam effects in asymmetric lepton-hadron colliders \cite{Mao}.


\section*{Acknowledgements}
It is a pleasure to express my gratitude to Prof. Igor Meshkov for his careful reading of the manuscript and for his useful comments and suggestions. Fruitful discussions on topics touched upon in the present article with Dr. Alexander Philippov are also gratefully acknowledged.


\appendix

\section{\label{sec:appendixB}Some Properties and Adjoint of the Frobenius-Perron Operator}

Combining Eqs. \eqref{FrobPerFinal} and \eqref{Rotation} the Frobenius-Perron operator \eqref{FrobPerOper} for the symplectic beam-beam twist map can be represented in a compact form as follows 
\begin{equation}
{\widehat{\mathbf{U}}}_k = {\widehat{\boldsymbol{\mathfrak{R}}}} {\left( - \omega_k \right)} \exp{\left\{ \lambda_k {\left[ \partial_q V_{3-k} {\left( q \right)} \right]} \partial_p \right\}}, \label{FroPerExpon}
\end{equation}
where ${\widehat{\boldsymbol{\mathfrak{R}}}} {\left( \alpha \right)}$ denotes the rotation of the canonical coordinates in phase space by an angle $\alpha$, specified by the orthogonal matrix \eqref{Rotation}. 

First, let us establish one of the most important properties of the Frobenius-Perron operator concerning the conservation of phase space volume. Consider the following integral 
\begin{equation}
\int {\rm d} q {\rm d} p {\widehat{\mathbf{U}}}_k f_k {\left( q, p \right)} = \int {\rm d} q {\rm d} p f_k {\left[ Q, P + \lambda_k V_{3-k}^{\prime} {\left( Q  \right)} \right]}. \label{ConservPSVol}
\end{equation}
Replacing the integration variables ${\left( q, p \right)} \longrightarrow {\left( Q, P \right)}$ on the right-hand side of the above equation and taking into account the fact that ${\rm d} q {\rm d} p = {\rm d} Q {\rm d} P$ under orthogonal transformation, we rewrite the right-hand side as  
\begin{equation} 
\int {\rm d} Q {\rm d} P \sum \limits_{m=0}^{\infty} {\frac {\lambda_k^m} {m!}} {\left[ \partial_Q V_{3-k} {\left( Q \right)} \right]}^m \partial_P^m f_k {\left( Q, P \right)} = \int {\rm d} Q {\rm d} P f_k {\left( Q, P \right)} = V_{ps}. \label{ConserPSVol} 
\end{equation}
In the left-hand side of the above equation, successive integration by parts has been performed and the independence of the beam-beam potential on the momentum variable $P$ has been taken into account.

The operator ${\widehat{\mathbf{U}}}_k^{\dagger}$ adjoint to the Frobenius-Perron operator ${\widehat{\mathbf{U}}}_k$ acting in phase space is defined by \cite{Arfken} 
\begin{equation}
\int {\rm d} q {\rm d} p g_k {\left( q, p \right)} {\widehat{\mathbf{U}}}_k f_k {\left( q, p \right)} = \int {\rm d} q {\rm d} p f_k {\left( q, p \right)} {\widehat{\mathbf{U}}}_k^{\dagger} g_k {\left( q, p \right)}. \label{DefAdjo}
\end{equation}
We transform the left-hand side of Eq \eqref{DefAdjo} as follows 
\begin{equation}
\begin{split}
\int {\rm d} q {\rm d} p g_k {\left( q, p \right)} {\widehat{\mathbf{U}}}_k f_k {\left( q, p \right)} = \int {\rm d} q {\rm d} p g_k {\left( q, p \right)} \exp{\left\{ \lambda_k {\left[ \partial_Q V_{3-k} {\left( Q \right)} \right]} \partial_P \right\}} f_k {\left( Q, P \right)} 
\\ 
= \int {\rm d} Q {\rm d} P {\widehat{\boldsymbol{\mathfrak{R}}}} {\left( \omega_k \right)} g_k {\left( Q, P \right)} \exp{\left\{ \lambda_k {\left[ \partial_Q V_{3-k} {\left( Q \right)} \right]} \partial_P \right\}} f_k {\left( Q, P \right)} 
\\ 
= \int {\rm d} Q {\rm d} P f_k {\left( Q, P \right)} \exp{\left\{ - \lambda_k {\left[ \partial_Q V_{3-k} {\left( Q \right)} \right]} \partial_P \right\}} 
{\widehat{\boldsymbol{\mathfrak{R}}}} {\left( \omega_k \right)} g_k {\left( Q, P \right)}. \label{Adjoint} 
\end{split}
\end{equation}
From all of the above, it follows that the adjoint operator can be written in the form
\begin{equation}
{\widehat{\mathbf{U}}}_k^{\dagger} = \exp{\left\{ - \lambda_k {\left[ \partial_q V_{3-k} {\left( q \right)} \right]} \partial_p \right\}} 
{\widehat{\boldsymbol{\mathfrak{R}}}} {\left( \omega_k \right)}. \label{Koopman}
\end{equation}
The adjoint operator to the Frobenius-Perron operator is also known as the Koopman operator.


\section{\label{sec:appendixA}Derivation of the Amplitude Equation for a Generic Potential. Non-resonant Case}

To be as general as possible, consider an arbitrary potential $U {\left( a, J \right)}$ written in angle-action variables as 
\begin{equation}
U {\left( a, J \right)} = V_0 {\left( J \right)} + V {\left( a, J \right)}. \label{ArbPotVN}
\end{equation}
Respectively, the Liouvillian operator can be written as 
\begin{equation}
{\widehat{\mathbf{L}}}_U = {\widehat{\mathbf{L}}}_0 + {\widehat{\mathbf{L}}}, \label{LiouvOperArb}
\end{equation}
where 
\begin{equation}
{\widehat{\mathbf{L}}}_0 = - \omega_u {\left( J \right)} \partial_a, \qquad \quad  {\widehat{\mathbf{L}}} = {\left( \partial_a V \right)} \partial_J - {\left( \partial_J V \right)} \partial_a, \label{LioOper0P}
\end{equation}
and 
\begin{equation}
\omega_u {\left( J \right)} = \partial_J V_0. \label{NonlinFreq}
\end{equation}
The Frobenius-Perron operator, which will be the subject of renormalization group reduction in this Appendix can be written as 
\begin{equation}
f_{n+1} {\left( a + \omega, J \right)} = \exp {\left( \epsilon {\widehat{\mathbf{L}}}_U \right)} f_n {\left( a, J \right)}. \label{FrobPerArb}
\end{equation}
In what follows, we adhere closely to Ref. \citenum{tzenovFP}. 

Initially, we consider the case, where the basic rotation frequency $\omega$ is away from nonlinear resonances driven by the potential $V$. Following the standard procedure of the renormalization group method \cite{tzenovBOOK}, we seek a solution to equation \eqref{FrobPerArb} by naive perturbation expansion 
\begin{equation}
f_n {\left( a, J \right)} = \sum \limits_{k=0}^{\infty} \epsilon^k f_n^{(k)} {\left( a, J \right)}, \label{FrobPerPert}
\end{equation}
where the unknown functions $f_n^{(k)} {\left( a, J \right)}$ should be determined order by order. 

\subsection{\label{sec:appAsecterm}Calculation of Secular Terms} 

\textbf{1. The zero-order equation} 
\begin{equation}
f_{n+1}^{(0)} {\left( a + \omega, J \right)} = f_n^{(0)} {\left( a, J \right)}, \label{FrobPer0Ord}
\end{equation}
possesses the obvious solution
\begin{equation}
f_n^{(0)} {\left( a, J \right)} = \exp {\left( - n \omega \partial_a \right)} F {\left( a, J \right)} = F {\left( a_n, J \right)}, \label{FroPer0Sol}
\end{equation}
where $a_n = a - n \omega$. To this end $F {\left( a, J \right)}$ is a generic function of its arguments, which will be the subject of the renormalization group reduction in the sequel.

\vspace{10pt} 

\textbf{2. The first-order equation} can be written as follows 
\begin{equation}
f_{n+1}^{(1)} {\left( a + \omega, J \right)} - f_n^{(1)} {\left( a, J \right)} = {\left( {\widehat{\mathbf{L}}}_0 + {\widehat{\mathbf{L}}} \right)} F {\left( a_n, J \right)}. \label{FrobPer1Ord}
\end{equation}
Note that since the above equation is linear, the right-hand side will give rise to two kinds of terms in the corresponding solution, one of them secular (proportional to the discrete "time" $n$), while the other one is regular, containing oscillation harmonics of the rotation frequency $\omega$ and the angle variable $a$. For this purpose, let us represent the solution of Eq. \eqref{FrobPer1Ord} as follows 
\begin{equation}
f_n^{(1)} {\left( a, J \right)} = \phi_n {\left( a, J \right)} + \psi_n {\left( a, J \right)}. \label{FrobPer1Rep}
\end{equation}
The first term (which will turn out to be secular) satisfies the equation 
\begin{equation}
\phi_{n+1} {\left( a + \omega, J \right)} - \phi_n {\left( a, J \right)} = {\widehat{\mathbf{L}}}_0 F {\left( a_n, J \right)}. \label{FrobPer1Ftm}
\end{equation}
It is straightforward to verify that the solution of the above equation is
\begin{equation}
\phi_n {\left( a, J \right)} = n {\widehat{\mathbf{L}}}_0 F {\left( a_n, J \right)}, \label{FroPer1SFT}
\end{equation}
The second term in the representation \eqref{FrobPer1Rep} satisfies the equation 
\begin{equation}
\psi_{n+1} {\left( a + \omega, J \right)} - \psi_n {\left( a, J \right)} = {\widehat{\mathbf{L}}} F {\left( a_n, J \right)}. \label{FrobPer1Stm}
\end{equation}
Since the angle-dependent part of potential $V {\left( a, J \right)}$, the arbitrary function $F {\left( a, J \right)}$ and the sought-for function $\psi_n {\left( a, J \right)}$ are periodic in the angle variable $a$, we can represent them as a Fourier series 
\begin{equation}
V {\left( a, J \right)} = \sum \limits_{m \neq 0} V_m {\left( J \right)} {\rm e}^{i m a}, \quad F {\left( a, J \right)} = \sum \limits_{s} F_s {\left( J \right)} {\rm e}^{i s a}, \label{FourRepr1}
\end{equation}
\begin{equation}
\psi_n {\left( a, J \right)} = \sum \limits_{k} G_k^{(n)} {\left( J \right)} {\rm e}^{i k a}. \label{FourRepr2}
\end{equation}
We substitute the above expansions into both sides of Eq. \eqref{FrobPer1Stm} and after equating similar harmonics, we obtain
\begin{equation}
G_k^{(n+1)} {\rm e}^{i k \omega} - G_k^{(n)} = \sum \limits_m {\left[ i m V_m F_{k-m}^{\prime} - i {\left( k - m \right)} V_m^{\prime} F_{k-m} \right]} {\rm e}^{-i {\left( k - m \right)} n \omega}. \label{FourBalan}
\end{equation}
Here, the primes indicate differentiation with respect to the action variable $J$. It is straightforward to verify that the solution of equation \eqref{FourBalan} has the form
\begin{equation}
G_k^{(n)} = \sum \limits_m {\frac {{\rm e}^{-i m \omega / 2}} {2i \sin {\left( m \omega / 2 \right)}}} {\left[ i m V_m F_{k-m}^{\prime} - i {\left( k - m \right)} V_m^{\prime} F_{k-m} \right]} {\rm e}^{-i {\left( k - m \right)} n \omega}. \label{FrobPer1SoG}
\end{equation}
Substituting back expression \eqref{FrobPer1SoG} into the expansion \eqref{FourRepr2} for the function $\psi_n$ and rearranging terms, we obtain
\begin{equation}
\psi_n = \sum \limits_{m,s} {\frac {{\rm e}^{i m {\left( a - \omega / 2 \right)}}} {2i \sin {\left( m \omega / 2 \right)}}} {\left[ i m V_m F_s^{\prime} - i s V_m^{\prime} F_s \right]} {\rm e}^{i s {\left( a - n \omega \right)}}. \label{FrobPer1Sopsi}
\end{equation}
The expression \eqref{FrobPer1Sopsi} for $\psi_n$ can be converted to a closed form such that the first-order solution to the first-order equation \eqref{FrobPer1Ord} reads as 
\begin{equation}
f_n^{(1)} {\left( a, J \right)} = {\left( n {\widehat{\mathbf{L}}}_0 + {\widehat{\boldsymbol{\mathcal{L}}}}_{\omega} \right)} F {\left( a_n, J \right)}, \label{FroPer1Sol}
\end{equation}
where 
\begin{equation}
{\widehat{\boldsymbol{\mathcal{L}}}}_{\omega} = {\left( \partial_a V_{\omega} \right)} \partial_J - {\left( \partial_J V_{\omega} \right)} \partial_a. \label{FroPer1Lom}
\end{equation}
Furthermore, the potential $V_{\omega}$ is defined according to the expression 
\begin{equation}
V_{\omega} {\left( a, J \right)} = {\mathcal{V}}_1 {\left( a - {\frac {\omega} {2}}, J \right)}, \qquad \qquad {\mathcal{V}}_1 {\left( a, J \right)} = \sum \limits_{m \neq 0} {\frac {V_m {\left( J \right)} {\rm e}^{i m a}} {2i \sin {\left( m \omega / 2 \right)}}}. \label{FroPer1Vom}
\end{equation}

\vspace{10pt} 

\textbf{3. The second-order equation} is 
\begin{equation} 
f_{n+1}^{(2)} {\left( a + \omega, J \right)} - f_n^{(2)} {\left( a, J \right)} = {\left( {\widehat{\mathbf{L}}}_0 + {\widehat{\mathbf{L}}} \right)} f_n^{(1)} {\left( a, J \right)} + {\frac {1} {2}} {\left( {\widehat{\mathbf{L}}}_0 + {\widehat{\mathbf{L}}} \right)}^2 F {\left( a_n, J \right)}. \label{FrobPer2Ord} 
\end{equation}
Since we are interested in the secular solution of Eq. \eqref{FrobPer2Ord}, we retain on its right-hand side only terms that would yield a secular contribution. Thus, the second-order equation giving rise to a secular solution can be written as 
\begin{equation} 
f_{n+1}^{(2)} {\left( a + \omega, J \right)} - f_n^{(2)} {\left( a, J \right)} = {\left[ {\left( n + {\frac {1} {2}} \right)} {\widehat{\mathbf{L}}}_0^2 + n {\widehat{\mathbf{L}}} {\widehat{\mathbf{L}}}_0 + \Omega {\left( \omega, J \right)} \partial_a \right]} F {\left( a_n, J \right)}. \label{FrobPer2OSC} 
\end{equation}
where 
\begin{equation}
\Omega {\left( \omega, J \right)} = \sum \limits_{m=1}^{\infty} m \cot {\left( {\frac {m \omega} {2}} \right)} \partial_J {\left( V_m \partial_J V_m \right)}. \label{FroPerNFreq}
\end{equation}
Note that the last operator $\Omega {\left( \omega, J \right)} \partial_a$ on the right-hand side of Eq. \eqref{FrobPer2OSC} is the angle-independent part of the sum ${\widehat{\mathbf{L}}} {\widehat{\boldsymbol{\mathcal{L}}}}_{\omega} + {\widehat{\mathbf{L}}}^2 / 2$. 

The first and last terms on the right-hand side of Eq. \eqref{FrobPer2OSC} can be treated in a way analogous to the treatment of Eq. \eqref{FrobPer1Ftm}. Consider the solution of the equation
\begin{equation}
\Psi_{n+1} {\left( a + \omega, J \right)} - \Psi_n {\left( a, J \right)} = n {\widehat{\mathbf{L}}} {\widehat{\mathbf{L}}}_0 F {\left( a_n, J \right)}. \label{FrobPer2Stm}
\end{equation}
Using the representations \eqref{FourRepr1} and \eqref{FourRepr2}, we can write the solution of the above equation as 
\begin{equation}
\Psi_n {\left( a, J \right)} = \sum \limits_{k} {\mathcal{G}}_k^{(n)} {\left( J \right)} {\rm e}^{i k a}. \label{FourRepr3}
\end{equation}
It can be verified by direct substitution that the functions ${\mathcal{G}}^{(n)}_k {\left( J \right)}$ are given by the expression
\begin{equation}
{\mathcal{G}}^{(n)}_k = \sum \limits_m {\left( n A_{km} + B_{km} \right)} {\rm e}^{-i {\left( k - m \right)} n \omega}, \label{FrobPer2SolG}
\end{equation}
where 
\begin{equation}
A_{km} = {\frac {{\rm e}^{-i m \omega / 2}} {2i \sin {\left( m \omega / 2 \right)}}} {\left[ i m V_m W_{k-m}^{\prime} - i {\left( k - m \right)} V_m^{\prime} W_{k-m} \right]}, \label{FrobPer2Akm}
\end{equation}
\begin{equation}
B_{km} = {\frac {1} {4 \sin^2 {\left( m \omega / 2 \right)}}} {\left[ i m V_m W_{k-m}^{\prime} - i {\left( k - m \right)} V_m^{\prime} W_{k-m} \right]}, \label{FrobPer2Bkm}
\end{equation}
and the new function $W {\left( a, J \right)} = {\widehat{\mathbf{L}}}_0 F {\left( a, J \right)}$ have been introduced. Based on the obvious parallel between Eq. \eqref{FourRepr3} and Eq. \eqref{FrobPer1Sopsi}, for the secular solution of the second-order equation \eqref{FrobPer2OSC} we obtain 
\begin{equation}
f_n^{(2)} {\left( a, J \right)} = {\left[ {\frac {n^2} {2}} {\widehat{\mathbf{L}}}_0^2 + n {\widehat{\boldsymbol{\mathcal{L}}}}_{\omega} {\widehat{\mathbf{L}}}_0 + n \Omega {\left( \omega, J \right)} \partial_a \right]} F {\left( a_n, J \right)}. \label{FroPer2Sol}
\end{equation}

\subsection{\label{sec:appAampeq}Derivation of the Amplitude Equation} 

To remove secular terms (proportional to $n$ and $n^2$) in the first-order \eqref{FroPer1Sol} and second-order \eqref{FroPer2Sol} solutions, we define a renormalization group transformation $F {\left( a, J \right)} \longrightarrow {\widetilde{F}} {\left( a, J; n \right)}$ by collecting all terms proportional to $F {\left( a_n, J \right)}$ 
\begin{equation}
{\widetilde{F}} {\left( a_n, J; n \right)} = {\left[ 1 + \epsilon n {\widehat{\mathbf{L}}}_0 + \epsilon^2 {\left( {\frac {n^2} {2}} {\widehat{\mathbf{L}}}_0^2 + n \Omega \partial_a \right)} \right]} F {\left( a_n, J \right)}. \label{FroPerRGtra}
\end{equation}
Solving perturbatively the above equation for $F {\left( a_n, J \right)}$ in terms of ${\widetilde{F}} {\left( a_n, J; n \right)}$, we obtain the following 
\begin{equation}
F {\left( a_n, J \right)} = {\left( 1 - \epsilon n {\widehat{\mathbf{L}}}_0 + \dots \right)} {\widetilde{F}} {\left( a_n, J; n \right)}. \label{FroPerRGinv}
\end{equation}
Following Refs. \citenum{tzenovBOOK,goto1,goto2} we define a discrete version of the renormalization group amplitude equation by considering the difference
\begin{equation}
{\widetilde{F}} {\left( a_n, J; n + 1 \right)} - {\widetilde{F}} {\left( a_n, J; n \right)} = {\left\{ \epsilon {\widehat{\mathbf{L}}}_0 + \epsilon^2 {\left[ {\left( n + {\frac {1} {2}} \right)} {\widehat{\mathbf{L}}}_0^2 + \Omega \partial_a \right]} \right\}} F {\left( a_n, J \right)}. \label{FroPerRGeq1}
\end{equation}
Substituting the expression for $F {\left( a_n, J \right)}$ in terms of ${\widetilde{F}} {\left( a_n, J; n \right)}$ from Eq. \eqref{FroPerRGinv}, we can eliminate secular terms up to $O {\left( \epsilon^2 \right)}$. The result is as follows  
\begin{equation}
{\widetilde{F}} {\left( a_n, J; n + 1 \right)} - {\widetilde{F}} {\left( a_n, J; n \right)} {\left[ \epsilon {\widehat{\mathbf{L}}}_0 + \epsilon^2 {\left( {\frac {1} {2}} {\widehat{\mathbf{L}}}_0^2 + \Omega \partial_a \right)} \right]} {\widetilde{F}} {\left( a_n, J; n \right)}. \label{FroPerRGeq}
\end{equation}
Equation \eqref{FroPerRGeq} is the sought-for renormalization group amplitude equation. It describes the evolution of the distribution function on slower time scales in addition to the fast regular oscillations with the fundamental rotation frequency $\omega$.

An important remark is in order at this point. Note that once the renormalization transformation has been performed, the second term (which is proportional to $n$ and, therefore, secular) in the second order solution \eqref{FroPer2Sol} is eliminated automatically. To see this, combine it with the second (non-secular) term in the first-order solution \eqref{FroPer1Sol}. Thus, we obtain 
\begin{equation} 
\epsilon {\widehat{\boldsymbol{\mathcal{L}}}}_{\omega} F + \epsilon^2 n {\widehat{\boldsymbol{\mathcal{L}}}}_{\omega} {\widehat{\mathbf{L}}}_0 F = \epsilon {\widehat{\boldsymbol{\mathcal{L}}}}_{\omega} {\left( 1 - \epsilon n {\widehat{\mathbf{L}}}_0 \right)} {\widetilde{F}} {\left( n \right)} + \epsilon^2 n {\widehat{\boldsymbol{\mathcal{L}}}}_{\omega} {\widehat{\mathbf{L}}}_0 {\widetilde{F}} {\left( n \right)} = \epsilon {\widehat{\boldsymbol{\mathcal{L}}}}_{\omega} {\widetilde{F}} {\left( n \right)}. \label{FroPerRGren}
\end{equation}

To first order in the perturbation parameter $\epsilon$ the renormalized solution to Eq. \eqref{FrobPerArb} can be written as
\begin{equation}
f_n {\left( a, J \right)} = {\left( 1 + \epsilon {\widehat{\boldsymbol{\mathcal{L}}}}_{\omega} \right)} {\widetilde{F}} {\left( a_n, J; n \right)}, \label{FroPerRGresol}
\end{equation}
where the renormalized amplitude ${\widetilde{F}} {\left( a_n, J; n \right)}$ satisfies the renormalization group amplitude equation \eqref{FroPerRGeq}. In the continuous limit Eq. \eqref{FroPerRGeq} acquires the form 
\begin{equation}
{\frac {\partial {\widetilde{F}} {\left( a_n, J; n \right)}} {\partial n}} = {\left[ \epsilon {\widehat{\mathbf{L}}}_0 + \epsilon^2 {\left( {\frac {1} {2}} {\widehat{\mathbf{L}}}_0^2 + \Omega \partial_a \right)} \right]} {\widetilde{F}} {\left( a_n, J; n \right)}. \label{FroPerRGeqcl}
\end{equation}
The above is a Fokker-Planck equation with the Fokker-Planck operator acting only on the angle variable.

In order to modify the renormalization reduction procedure of the Frobenius-Perron operator just described into an approach suitable for our purposes here, an important addition is necessary to be worked out at this point. Since the beam-beam potential functionally depends on the distribution function, it itself represents a perturbation series in $\epsilon$ in the sense of an order of magnitude. Thus, the representation \eqref{ArbPotVN} must be replaced by 
\begin{equation}
U {\left( a, J \right)} = V_0 {\left( J \right)} + V {\left( a, J \right)} + \epsilon {\left[ W_0 {\left( J \right)} + W {\left( a, J \right)} \right]} + \dots. \label{ArbPotVNW} 
\end{equation}
Respectively, the Liouvillian operator can be written as 
\begin{equation}
{\widehat{\mathbf{L}}}_U = {\widehat{\mathbf{L}}}_0 + {\widehat{\mathbf{L}}} + \epsilon {\left( {\widehat{\mathbf{M}}}_0 + {\widehat{\mathbf{M}}} \right)}, \label{LiouvOperArbW}
\end{equation}
where 
\begin{equation}
{\widehat{\mathbf{M}}}_0 = - \omega_w {\left( J \right)} \partial_a, \qquad \quad  {\widehat{\mathbf{M}}} = {\left( \partial_a W \right)} \partial_J - {\left( \partial_J W \right)} \partial_a, \label{LioOper0PW}
\end{equation}
and 
\begin{equation}
\omega_w {\left( J \right)} = \partial_J W_0. \label{NonlinFreqW}
\end{equation}
The result of adding the first-order potentials $W_0$ and $W$ (depending on the first-order distribution function) is the appearance of an additional term $\epsilon^2 {\widehat{\mathbf{M}}}_0 {\widetilde{F}} {\left( a_n, J; n \right)}$ on the right-hand side of the renormalization group amplitude equation \eqref{FroPerRGeq}. All this term does is introduce a higher-order correction to the incoherent tune shift without changing the character of the Fokker-Planck operator in the amplitude equation, as this operator continues to act in the subspace of the angle variables alone.





\section*{Conflicts of Interest}

The author declares no conflicts of interest.



\bibliographystyle{nsr}
\bibliography{BEAM_BEAM}



\end{document}